# Multi-Graph Convolutional Network for Short-Term Passenger Flow Forecasting in Urban Rail Transit

Jinlei Zhang [1], Feng Chen [1, 2, 3*], Yinan Guo[4], Xiaohong Li[1]

[1] School of Civil Engineering, Beijing Jiaotong University, No. 3 Shangyuancun, Haidian District, Beijing 100044, China

[2] School of Highway, Chang'an University, Middle-section of Nan'er Huan Road, Xi'an 710064, China

[3] Beijing Engineering and Technology Research Centre of Rail Transit Line Safety and Disaster Prevention, No.3 Shangyuancun, Haidian District, Beijing 100044, China

[4] Information School, University of Washington, Seattle, WA, USA

* Correspondence: fengchen@bjtu.edu.cn

**Abstract:** Short-term passenger flow forecasting is a crucial task for urban rail transit operations. Emerging deep-learning technologies have become effective methods used to overcome this problem. In this study, we propose a deep-learning architecture called Conv-GCN that combines a graph convolutional network (GCN) and a three-dimensional (3D) convolutional neural network (3D CNN). First, we introduce a multi-graph GCN to deal with three inflow and outflow patterns (recent, daily, and weekly) separately. Multi-graph GCN networks can capture spatiotemporal correlations and topological information within the entire network. A 3D CNN is then applied to deeply integrate the inflow and outflow information. High-level spatiotemporal features between different inflow and outflow patterns and between stations that are nearby and far away can be extracted by 3D CNN. Finally, a fully connected layer is used to output results. The Conv-GCN model is evaluated on smart card data of the Beijing subway under the time interval of 10 min, 15 min, and 30 min. Results show that this model yields the best performance compared with seven other models. In terms of the root-mean-square errors, the performances under three time intervals have been improved by 9.402%, 7.756%, and 9.256%, respectively. This study can provide critical insights for subway operators to optimise urban rail transit operations.

## 1. Introduction

Short-term passenger flow forecasting (STPFF) is of critical importance in urban rail transit (URT). Passengers can schedule their trips in advance by utilising forecasting results. Operators can take immediate measures to avoid traffic congestions. However, this is a challenging task for a citywide prediction because it is easily affected by many factors, such as spatiotemporal dependencies, topological information, incidents, and weather conditions. Models for STPFF range from statistics-based models, such as historical average and autoregressive integrated moving average (ARIMA) [1, 2], to machine-learning-based models, such as neural networks, and support vector machines [3-5]. Recently, deep-learning-based models have been widely introduced to tackle this problem and have been proved to have great advantages than previous models. For example, they have favourable prediction precisions and can meet real-time requirements. We can also use one model to make predictions in a citywide network [6, 7]. Deep-learning-based models can be summarized as follows.

In the early stage, some models involve the recurrent-neural-network (RNN)-based models. Ma et al. [8] introduced the long short-term memory (LSTM) network for traffic speed prediction for the first time. Similarly, Yang et al. [9] also applied the LSTM network. They utilised the enhanced long-term features to improve prediction precision. Fu et al. [10] applied gated recurrent unit (GRU) to perform traffic flow prediction for the first time. However, they only considered the temporal dependencies. Zheng et al. [11] proposed an LSTM architecture via a two-dimensional network which can consider spatiotemporal correlations for short-term traffic forecasts. Zhang et al. [7] built a cluster-based LSTM model to conduct STPFF in URT that can be used when the available data are limited. Generally, RNN-based models cannot consider spatial correlations in a citywide network. Moreover, it will take a longer training time because parallel computing cannot be utilised during the training processes.

After the RNN was applied to traffic prediction, researchers found the promising performance of convolutional-neural-network (CNN), which can extract spatial dependencies even when stations are far away from each other. CNN-based models always treat passenger flows as images to allow the execution of convolution operations [12]. The residual network (ResNet) [13] is a typical framework using skip-connection between CNN layers. It has been proved to be effective in STPFF, such as spatiotemporal ResNet models [14, 15]. However, CNN-based models can only be used for Euclidean data. All traffic data that are actually Non-Euclidean must be transformed into structural data with a fixed form so that they can be input in CNN-based models. Therefore, some structural information in the network will be lost during pre-processing.

In recent years, the graph-convolutional-network (GCN) becomes prevailing because of its better performance in traffic prediction [16]. These models can capture spatiotemporal correlations and topological information between stations or areas. The structural information of Non-Euclidean data can be fully utilised. Moreover, they are associated with faster training speeds and fewer parameters than RNN- and CNN-based models. Some models considered recent, daily, and weekly patterns during the graph convolutional process [6, 17, 18]. Recent studies constructed multi-graph networks to capture several types of adjacent information, such as proximity, connectivity, and functionality, to improve precision [19, 20]. Yu et al. [21] introduced the STGCN model that achieved an increased training speed with fewer parameters, and performed better than many other models. Some researchers built an extended GCN model considering area-wide spatiotemporal dependencies [22]. The GCN-based models, however, generally use one to four GCN layers. They cannot go as deep as CNN-based models [23]. Therefore, some deep spatial correlations cannot be effectively captured.

To effectively extract spatiotemporal features, some studies built complex architectures involving RNN, CNN, or GCN, etc. For example, some studies integrated GCN and LSTM or GRU to make traffic predictions [24-27]. Park et al. [28] used the transformer model [29] and self-attention mechanism with an encoder-decoder architecture. Hao et al. [30] built a sequence to sequence architecture with an attention mechanism to conduct STPFF in a metro system. Zhang et al. [31] designed an architecture that included the attention mechanism, GCN, and sequence-to-sequence model to conduct multistep speed prediction. Guo et al. [32] combined an autoencoder network with kernel ridge regression and Gaussian process regression to make short-term prediction in URT. Jia et al. [33] integrated the LSTM and stacked auto-encoders for predicting short-term passenger flows of each station in a subway network simultaneously. After ConvLSTM was firstly introduced [34], CNN and LSTM were often integrated to perform traffic predictions [35-37]. Recently, generative adversarial networks have begun to attract researchers' attention and have been applied to traffic time estimation [38]. Generally, these deep-learning architectures are so complicated that it is difficult to reproduce or transplant. Moreover, they consume tremendous computing resources and training time.

Overall, existing models present several drawbacks. First, some models cannot capture spatial and temporal dependencies simultaneously. For example, the RNN-based models can only capture temporal correlations, while the spatial correlations are missed. Second, the overall topological information was neglected sometimes. Both RNN- and CNN-based models miss the topological information. Most existing GCN-based models only involve one graph that cannot thoroughly extract topological information. Third, two-dimensional (2D) CNN cannot organically integrate the inflow and outflow information. Only a few research studies have focused on the application of three-dimensional (3D) CNN that can effectively extract high-level features leveraging 3D filters. Finally, many models are so complicated that a lot of computing resources and time will be cost during the training process. Models are not the more complicated the better. Identifying methodologies to improve the prediction's precision using relatively simpler approaches with more efficient models is also important.

To overcome these shortcomings, we propose a deep-learning architecture called Conv-GCN based on a multi-graph GCN and a 3D convolutional network (3D CNN), which are relatively simple while more effective. The model is evaluated on smart card data obtained from the Beijing subway under three time intervals. The proposed model performance was always the best in all cases (compared with seven other models), thus showing strong robustness. The main contributions of this model are as follows.

(1) The Conv-GCN has a simpler while more efficient architecture because the two branches are based on GCN that can save more time to train. This proves that models are not the more complicated the better.

(2) The multi-graph GCN can capture the spatiotemporal and topological correlations in an entire network. Three inflow and outflow patterns (recent, daily, and weekly) are involved in this model. Two graph branches are utilised to extract inflow and outflow topological information.

(3) The 3D CNN can effectively integrate the inflow and outflow information via 3D filters. It can also capture high-level spatiotemporal features between three patterns of inflow and outflow, as well as between stations nearby and far away.

The remaining parts of this study are organised as follows. In section 2, we define the problem and present the model architecture. The multi-graph GCN and 3D CNN used in this model are also described. In section 3, the experimental details and results are discussed. The conclusion is outlined in section 4.

## 2. Methodology

In this section, the methodological framework is formulated. First, we define the problem that needs to be solved. Second, the model architecture is constructed. Third, one part of the architecture, namely, the multi-graph GCN, is summarised followed by the description of 3D CNN.

### 2.1. Problem definition

The goal of this study is to predict the tap-in ridership in the URT network using historical smart card data. The historical tap-in ridership can be extracted from smart card data and can be aggregated at different time intervals, such as 10 min, 15 min, and 30 min.

We define the URT network as a graph $G = (V, E, A)$, wherein $V$ are the vertices representing subway stations, $V = (V_1, V_2, V_3, \cdots V_n)$, wherein $n$ is the station number, and $E$ are the edges between stations. In addition, $A \in \mathbb{R}^{n \times n}$ is the adjacent matrix whose elements are all ones and zeros that indicate the existence (or absence) of a link between two stations. The feature matrix $F \in \mathbb{R}^{n \times m} = (X_t, X_{t-1}, X_{t-2}, \cdots X_{t-m+1})$, whereby $n$ is the station number that is ordered according to the subway line number, $m$ denotes the past several time intervals used to predict the ridership in the next time interval, and $X \in \mathbb{R}^{n \times 1}$ is the tap-in passenger flow vector in a specific time interval. Each time interval will generate a feature matrix. Therefore, the problem can be defined according to (1), whereby $f(\cdot)$ is the mapping function to be learnt using the proposed deep-learning architecture.

$$X_{t+1} = f(A; X_t, X_{t-1}, X_{t-2}, \cdots X_{t-m+1}) \qquad (1)$$

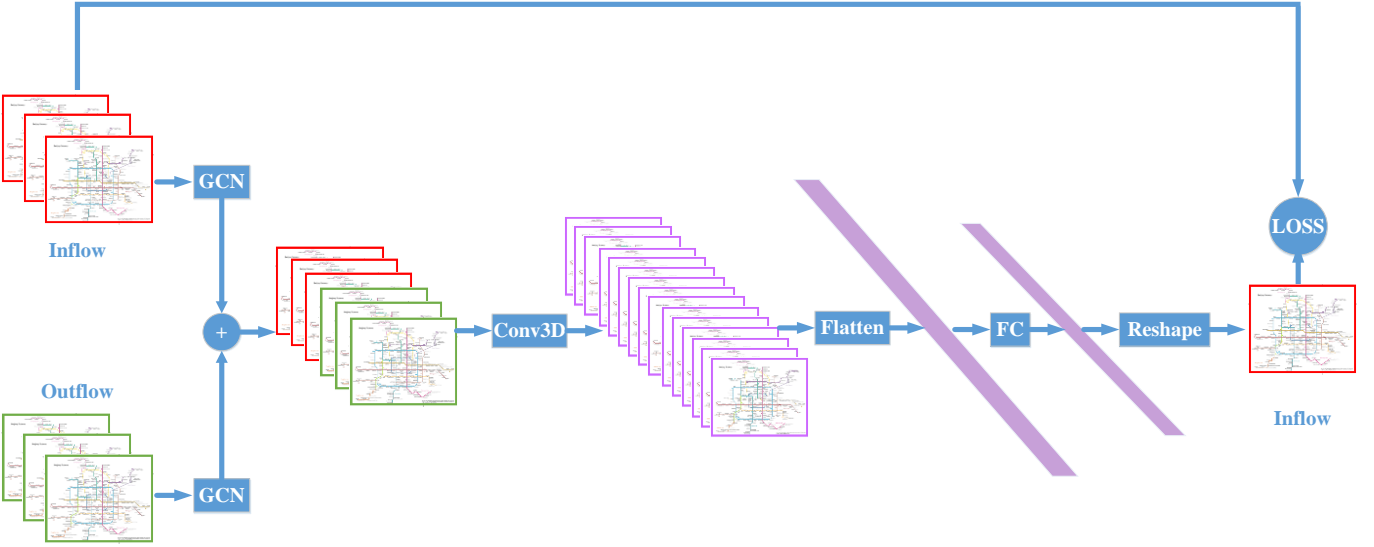

*Fig. 1* Model architecture

## 2.2. Model architecture

The Conv-GCN model architecture is shown in Fig. 1. First, the inflow graph (red graph) and outflow graph (green graph) are dealt with by the multi-graph GCN, which can easily capture spatiotemporal and topological information. In each graph, three travel patterns, the recent, daily, and weekly patterns, are taken into account to capture temporal correlations. The recent pattern denotes the passenger flow volumes in the last several time intervals. The daily and weekly patterns respectively denote the corresponding information in the same time interval on the last day and the same day of the past week. Subsequently, the outputs of GCN branches are concatenated together and are then input into the 3D CNN layer. The 3D CNN layer is used to deeply integrate the inflow and outflow information obtained from the GCN layer. The high-level spatiotemporal information can be extracted. Following the 3D CNN layer, the output is then flattened and input into the fully connected layer. The fully connected layer is used to reduce the dimensions of the flattening layer, as well as capture the non-linear relationship between the high-level features and the predicted results. The output of fully connected layer is finally reshaped into the target shape.

## 2.3. Introduction of Multi-graph GCN

The GCN plays a critical role in our proposed Conv-GCN model because of its powerful ability to capture spatiotemporal and topological information. In this study, we applied two graphs that dealt with the inflow and outflow. In recent years, GCN has received considerable attention. The GCN layer has experienced significant improvements owing to development documents in the spectral graph f [39], Chebyshev polynomial [40], and first-order filters [16]. The stack of the GCN layer with the first-order filter can achieve similar effects with the k-Chebyshev polynomial filter [21] while it can achieve a significantly higher training speed and prediction accuracy in most cases [16]. Therefore, we used the GCN proposed by Kipf et al. [15] as shown in (2).

$$X^{l+1} = \sigma\left(\widetilde{D}^{-\frac{1}{2}}\tilde{A}\widetilde{D}^{-\frac{1}{2}}X^l W^l + b\right), \tilde{A} = A + I \quad (2)$$

where $A \in \mathbb{R}^{n \times n}$ is the adjacent matrix, $I \in \mathbb{R}^{n \times n}$ is the identity matrix, $\widetilde{D} \in \mathbb{R}^{n \times n}$ is the diagonal node degree matrix of $\tilde{A}$, $X^l \in \mathbb{R}^{n \times m}$ is the feature matrix of the $l^{th}$ layer in which $m$ represents the time steps used to predict the ridership in the next time step, $W^l \in \mathbb{R}^{m \times k}$ is the weight matrix of the $l^{th}$ layer in which $k$ is the kernel number, namely the output feature number per node, $b \in \mathbb{R}^{k \times 1}$ is the bias vector, and $\sigma$ is the activation function.

The GCN diagram used in this study is shown in Fig. 2 and Fig. *3*. Let us take station *E* as an example. If there is only one subway line for the passenger flow prediction of station *E*, as shown in Fig. 2, the first GCN layer will capture the influences of the adjacent stations *D* and *F*. After stacking another GCN layer, the influences of the stations *C* and *D* will also be integrated in its prediction process. Similarly, in the subway network, if *E* is an interchange station of two subway lines, it will capture the influences of four adjacent stations in the first GCN layer, and the influences of stations adjacent to these four stations in the second GCN layer.

Moreover, we also considered three passenger flow patterns, namely, recent, daily, and weekly patterns. In each pattern, the same time steps were applied. This type of organisation method benefits the prediction precision from three aspects. First, it can capture the temporal correlation among different patterns. The spatial correlations between nearby stations and those which are far away can also be considered. Finally, the topological information between adjacent stations can also be fully utilised. Two branches deal with inflow and outflow separately.

Their results are concatenated together before being input into 3D CNN to deeply integrate inflow and outflow information as shown in Fig. 1 and (3).

$$V = V^{l+1}_{inflow} \oplus V^{l+1}_{outflow} \quad (3)$$

where *V* is the input for 3D CNN, $V^{l+1}_{inflow}$ and $V^{l+1}_{inflow}$ are the output of the GCN branches, "$\oplus$" means stacking the two output tensors together according to the concatenation axis without doing any other process.

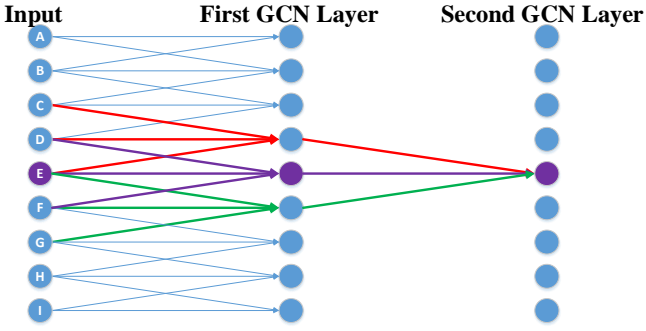

**Fig. 2** *Graph convolutional network (GCN) on one subway line*

### 2.4. Introduction of 3D Convolutional Network

In our Conv-GCN model, we used the 3D convolutional network (3D CNN) rather than a general 2D convolutional network (2D CNN), as shown in Fig. 4.

The 2D CNN can only extract features from local neighbourhood on the feature map from the previous layer. The values $v$ at position $(x, y, z)$ in the $j^{th}$ feature map of $i^{th}$ layer flow as follows [41].

$$v_{ij}^{xy} = ReLU\left(b_{ij} + \sum_m \sum_{p=0}^{P_i-1} \sum_{q=0}^{Q_i-1} w_{ijm}^{pq} v_{(i-1)m}^{(x+p)(y+q)}\right) \quad (4)$$

where ReLU is the activation function, $m$ is the feature map index in the $(i-1)^{th}$ layer, $w_{ijm}^{pq}$ is the $(p, q)^{th}$ value of the 2D kernel filter connected to $m^{th}$ feature map in the $(i-1)^{th}$ layer, $(P, Q)$ is the dimension of 2D filters.

The most important difference between 2D CNN and 3D CNN is the kernel dimension. The kernel dimension of 3D CNN is three dimensions. The values in 3D CNN flow as follows.

$$V^{l+1} = \text{Conv3D}(W \cdot V + b) \quad (5)$$

where $V$ is the input, $W$ is the 3D filters, $b$ is the bias, Conv3D represents the 3D convolution, $V^{l+1}$ is the high-level output feature. The values $v$ at position $(x, y, z)$ in the $j^{th}$ feature map of $i^{th}$ layer can be obtained by [41]:

$$v_{ij}^{xyz} = ReLU(b_{ij} + \sum_m \sum_{p=0}^{P_i-1} \sum_{q=0}^{Q_i-1} \sum_{r=0}^{R_i-1} w_{ijm}^{pqr} v_{(i-1)m}^{(x+p)(y+q)(z+r)}) \quad (6)$$

where ReLU is the activation function, $m$ is the feature map index in the $(i-1)^{th}$ layer, $w_{ijm}^{pqr}$ is the $(p, q, r)^{th}$ value of the 3D kernel connected to $m^{th}$ feature map in the $(i-1)^{th}$ layer, $(P, Q, R)$ is the dimension of 3D filters.

3D CNN has been extensively applied to computer vision, such as in medical image analysis, abnormal event detection, and human action recognition [41-43]. It has also been proven to be effective in learning spatiotemporal features [43] for several reasons. First, the 3D convolution kernel can effectively integrate information from different channels together. Second, 3D CNN can model spatiotemporal features in a better manner compared with 2D CNN because 3D convolution is performed spatiotemporally, while 2D convolution can only be performed spatially. Therefore, we applied a 3D CNN to aggregate the inflow and outflow information output by the GCN layer so that the outflow information can be fully utilised. High-level spatiotemporal features can be extracted among different patterns of inflow and outflow, as well as between nearby and far away stations.

The output of 3D CNN is flattened and input in a fully connected layer, as shown in Fig. 5 and (7).

$$V^{l+1} = f(W \cdot V + b) \quad (7)$$

where $f$ is the linear activation function.

The fully connected layer is used to reduce the data dimension, as well as capture the non-linear correlation between high-level features and outputs. We used only one fully connected layer to reduce the dimension of the flattening layer to the dimension we adopted. The output of the fully connected layer is finally reshaped into the final predicted results.

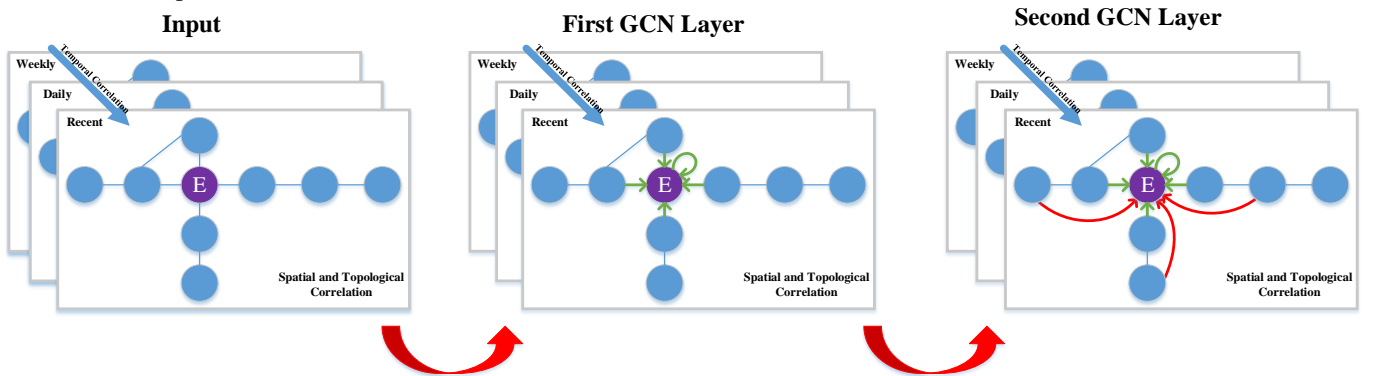

**Fig. 3** *GCN on subway network*

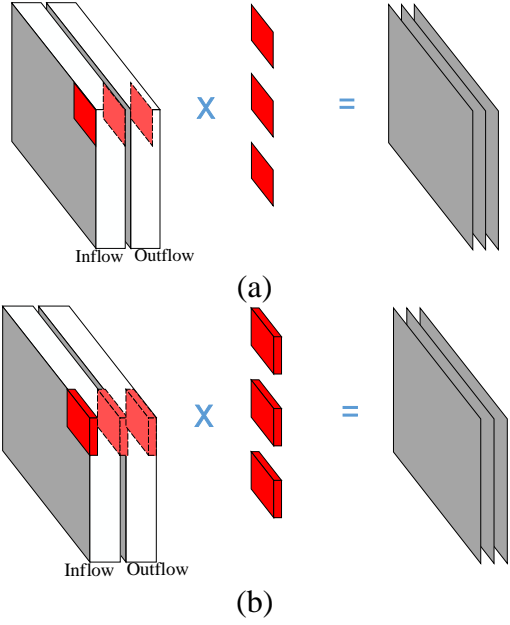

*Fig. 4 Diagram of two-dimensional (2D) convolutional neural network (CNN) (a) and three-dimensional (3D) CNN (b)*

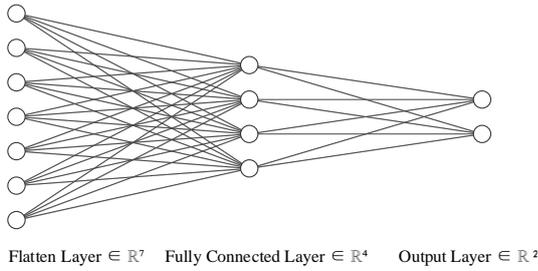

*Fig. 5 Diagram of fully connected layer*

## 3. Experiment

In this section, we will describe the dataset used in our study at first. The evaluation metrics are then presented. Several popular models are chosen as baseline models. The detailed experimental settings are also discussed. Finally, we analyse the predicted results.

### 3.1. Dataset description

The Conv-GCN model performance was evaluated with smart card data obtained from the Beijing Subway from February 29, 2016 to April 3, 2016. Each record contained the card number, tap-in time, tap-in station name, tap-out time, and tap-out station name as shown in Table 1. We gave each station a unique station index. The tap-in time and tap-out time from 05:00 to 23:00 were transformed into minutes. The records are integrated into specific time intervals, namely, 10 min, 15 min, and 30 min as shown in Table 2. We only used weekday data to evaluate the model. The data in the first four weeks were used to train the model and the remaining data in the last week were used to test the model. There were 276 stations in March 2016. Therefore, the adjacent matrix is $A \in \mathbb{R}^{276 \times 276}$. The passenger flow information, that is, the feature matrix, was incorporated in $F \in \mathbb{R}^{276 \times m}$. Each row of the feature matrix represented a subway station, and each column represented the ridership in the specific time interval as shown in Table 2. All data were normalised to the range (0, 1) with min-max scalers. The result evaluation was conducted after the predicted results were rescaled to their original scale range.

**Table 1** Original smart card data record

| Card number | Tap-in time | Tap-in station name | Tap-out time | Tap-out station name |
|---|---|---|---|---|
| 1987**** | 2016/3/5 08:33:00 | Xidan | 2016/3/5 09:21:41 | Shangdi |
| 2982**** | 2016/3/5 09:21:00 | Bagou | 2016/3/5 09:21:35 | Xiju |
| 3356**** | 2016/3/5 06:43:00 | Xidan | 2016/3/5 07:02:40 | dongdan |
| … | … | … | … | … |

**Table 2** Tap-in passenger flow series

| Station index | 05:00-05:15 | 05:15-05:30 | 05:30-05:45 | … | 22:45-23:00 |
|---|---|---|---|---|---|
| 1 | 30 | 55 | 77 | … | 22 |
| 2 | 15 | 42 | 58 | … | 11 |
| 3 | 18 | 37 | 49 | … | 19 |
| … | … | … | … | … | … |
| 276 | 23 | 47 | 62 | … | 16 |

### 3.2. Evaluation metrics and loss function

We chose the mean square error (MSE) as the loss function. Three indicators, the root mean square error (RMSE), mean absolute error (MAE), and weighted mean absolute percentage error (WMAPE), were used to evaluate model performances, as shown by (3)– (6).

$$Loss = MSE = \frac{1}{N}\sum_{i=1}^{N}(\tilde{X}_i - X_i)^2 \quad (3)$$

$$RMSE = \sqrt{\frac{1}{N}\sum_{i=1}^{N}(\tilde{X}_i - X_i)^2} \quad (4)$$

$$MAE = \frac{1}{N}\sum_{i=1}^{N}|\tilde{X}_i - X_i| \quad (5)$$

$$WMAPE = \sum_{i=1}^{N}(\frac{X_i}{\sum_{j=1}^{N}X_j}\left|\frac{\tilde{X}_i - X_i}{X_i}\right|) \quad (6)$$

where $\tilde{X}_i$ is the predicted value and $X_i$ is the actual value. $N$ denotes the total number of values that need to be predicted.

### 3.3. Comparison with state-of-the-art models

We compared the performances of the Conv-GCN model with the following models. All models were generated and executed on a desktop computer with an Intel i7-8700K processor (12M Cache, up to 4.7 GHz), 8 GB memory, and an NVIDIA GeForce GTX 1070 Ti.

**HA:** Historical average model. We used the average values in the last time step of three patterns to predict the value in the next time step [1].

- **ARIMA:** Autoregressive integrated moving average model. We used the Expert Modeller in the Statistical-Package-for-the-Social-Sciences (SPSS®IBM Corp., USA) to obtain the ARIMA results [44]. The Expert Modeller in SPSS can automatically give the best predicted results
- **LSTM:** LSTM was first applied to traffic field in 2015 [8]. We applied an LSTM model with two hidden layers and one fully connected layer. Each LSTM hidden layer has 100 neurons. The optimizer is Adam with learning rete as 0.001. The fully connected layer has one neuron. We use data of 276 stations to train one LSTM model. The inputs are the inflow series of 276 stations in the last 5 time steps. The outputs are the inflow series of 276 stations in the next time step.
- **2D CNN:** We applied a general 2D CNN model with two layers one fully connected layer [12]. The two hidden layers have 32 and 64 filters, respectively. The kernel size is 3*3. The fully connected layer has 276 neurons. The optimizer is Adam with learning rete as 0.001. The inputs are inflow and outflow of 276 stations from three patterns in the last 5 time steps. The outputs are the inflow of 276 stations in the next time steps.
- **ConvLSTM:** ConvLSTM was proposed by Shi et al. [34]. It achieved success in 2015. We also applied a ConvLSTM model with two hidden layers and one fully connected layer. Other parameters are the same with the 2D CNN model.
- **ST-ResNet:** This was proposed by Zhang et al. [15]. Herein, we only adopted three branches of this model and did not use weather data. The other parameters are similar with [15].
- **3D CNN:** We built a 3D CNN model. This model has the same parameter with the proposed Conv-GCN while without the two GCN branch.
- **ST-GCN:** We built an ST-GCN model. This model has the same parameter with the proposed Conv-GCN while without the 3D CNN layer.
- **ResLSTM:** A deep-learning architecture comprised GCN, ResNet, and attention LSTM. The parameters are the same with [44].

### 3.4. Experimental settings

We used Keras and TensorFlow to implement our model. We applied one GCN layer both for inflow and outflow, and one 3D CNN layer after the concatenation of two GCN branches. There were several hyperparameters, the time steps, batch size, kernel number of GCN layer, and filter number of Conv3D layer. To obtain the best parameter combination, we set time steps as indicated in [3, 17]. We also set the batch size (4, 8, 16, 32, 64, 128, 256), filter number of the Conv3D layer (1, 2, 4, 8, 16, 32, 64), and the kernel number of the GCN layer (6, 9, 12, 15, 18, 21, 24) to different values. During parameter tuning, we used the **control variate method**. This means that we maintained the other three parameters unchanged during the tuning process of a single parameter until we found the best combination of the set of four parameters. For example, we first chose randomly a combination of the four parameters. The other three parameters were then unchanged during the tuning of the time step parameter. When the optimal time step was found, we maintained its value unchanged and began to tune the other three parameters. The testing results are shown in Fig. 6. According to the variation of RMSE and MAE, the time step, batch size, filter number, and kernel number, were set as 10, 64, 16, and 15, respectively. We respectively used 30, 20, and 10 time steps for the time intervals of 10 min, 15 min, and 30 min.

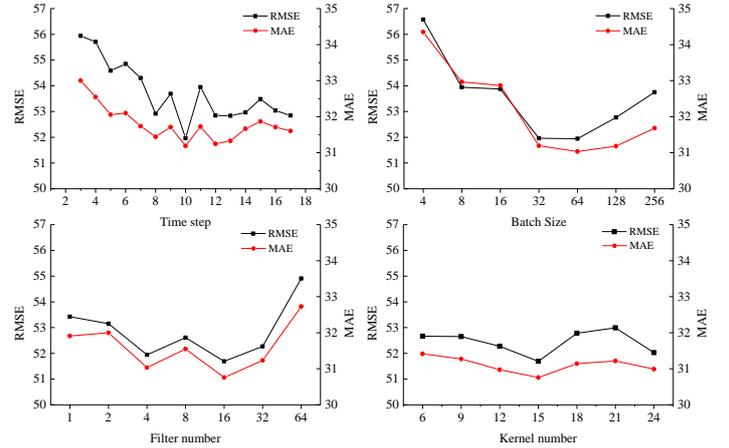

***Fig. 6** Testing results for chosen hyperparameters*

### 3.5. Result analyses

The results are shown in Table 3 and Fig. 7. As indicated, the best performance is associated with Conv-GCN, whichever the time interval is.

Conventional HA and ARIMA performs the worst no matter in short-term or long-term scenarios. The reason is that these two models can only capture limited temporal correlations. The important spatial and topological information is also missed during modeling. Because the LSTM can capture more temporal correlations and 2D CNN can capture more spatial correlations, the LSTM and 2D CNN performed better than conventional models. As it can be observed, complex deep-learning architectures like ConvLSTM, ST-ResNet, ResLSTM, and ST-GCN yielded more favourable results than single models in most cases. That is mainly because the spatiotemporal features can be extracted simultaneously in these models. It is worthy to mention that the 3D CNN showed promising results, which benefits greatly from the 3D filters. The proposed Conv-GCN yielded the best precision in all cases, and exhibited strong robustness because of the ingenious structure.

In terms of RMSE, the significant improvements compared with the best (existing) models were 9.402%, 7.756%, and 9.256% for the three time intervals. As for MAE, the respective improvements were 6.692%, 4.836%, and 5.602%. The corresponding improvements for WMAPE were 3.946%, 1.627%, and 2.804%.

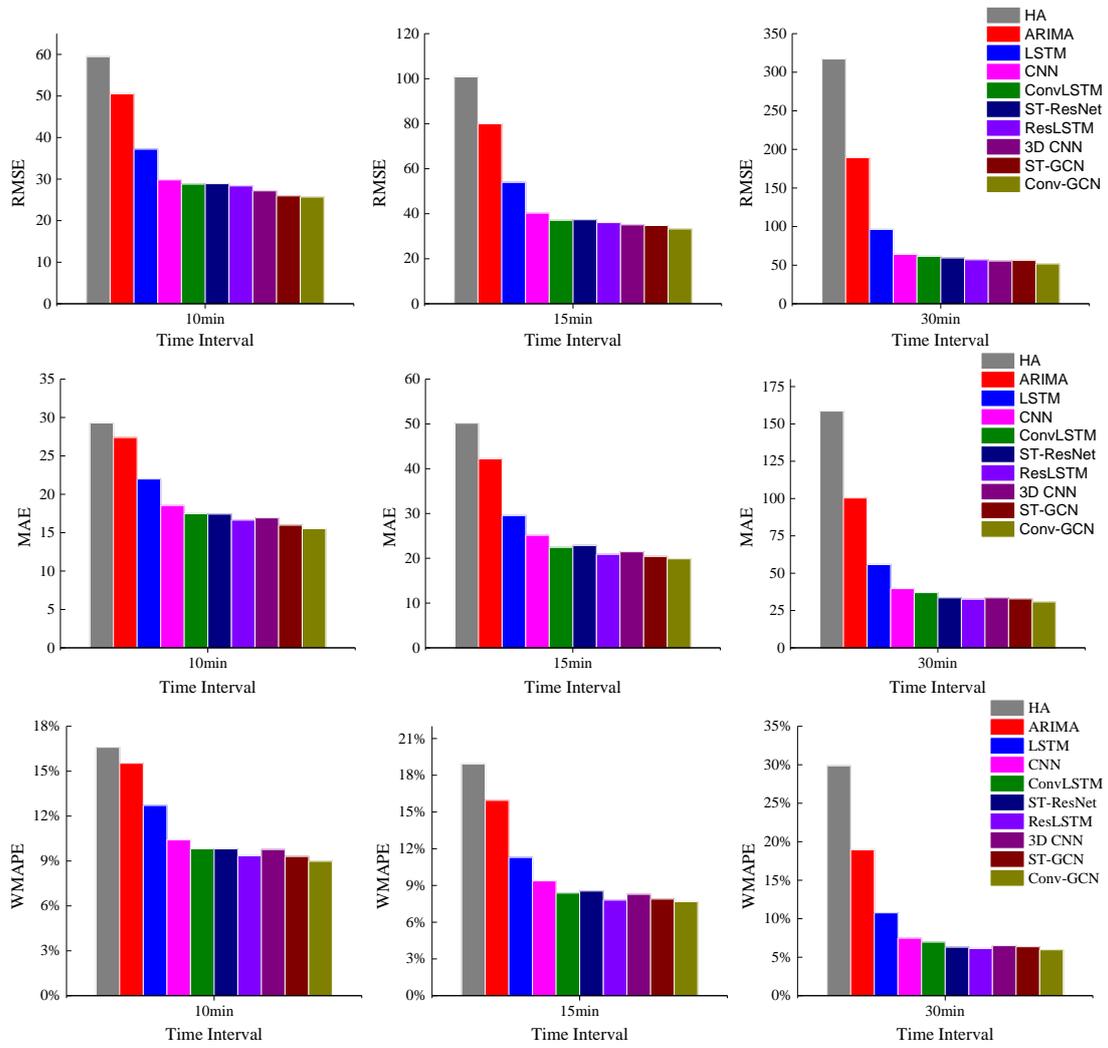
*Fig. 7 Model performance comparison for different time intervals*

**Table 3** Model performance comparison for different time intervals

| Time Interval | 10 min | | | 15 min | | | 30 min | | |
|---|---|---|---|---|---|---|---|---|---|
| | RMSE | MAE | WMAPE | RMSE | MAE | WMAPE | RMSE | MAE | WMAPE |
| HA | 59.4652 | 29.2990 | 16.60% | 100.8358 | 50.1607 | 18.93% | 317.1108 | 158.6099 | 29.88% |
| ARIMA | 50.5436 | 27.3968 | 15.53% | 79.9580 | 42.2139 | 15.95% | 189.3329 | 100.3590 | 18.95% |
| LSTM | 37.1903 | 21.9925 | 12.71% | 53.9216 | 29.5340 | 11.29% | 96.3534 | 55.8265 | 10.76% |
| 2D CNN | 29.8125 | 18.5460 | 10.413% | 40.2673 | 25.1231 | 9.375% | 64.0458 | 39.6867 | 7.472% |
| ConvLSTM | 28.7943 | 17.4780 | 9.814% | 37.0923 | 22.4236 | 8.380% | 61.4978 | 36.9768 | 6.962% |
| ST-ResNet | 28.8943 | 17.4224 | 9.812% | 37.3432 | 22.8570 | 8.545% | 59.3686 | 33.5018 | 6.309% |
| ResLSTM | 28.3661 | 16.6318 | 9.352% | 36.0444 | 20.8783 | 7.805% | 56.9649 | 32.5819 | 6.134% |
| 3D CNN | 27.1777 | 16.9238 | 9.768% | 35.0790 | 21.4495 | 8.281% | 55.4101 | 33.5078 | 6.494% |
| ST-GCN | 25.9634 | 15.9790 | 9.305% | 34.7628 | 20.4068 | 7.890% | 56.3030 | 32.8163 | 6.362% |
| Conv-GCN | **25.6992** | **15.5188** | **8.983%** | **33.2488** | **19.8687** | **7.678%** | **51.6925** | **30.7568** | **5.962%** |
| Improvement | **9.402%** | **6.692%** | **3.946%** | **7.756%** | **4.836%** | **1.627%** | **9.256%** | **5.602%** | **2.804%** |

## 4. Conclusion

This study proposed a deep-learning architecture called Conv-GCN to conduct STPFF in URT. The Conv-GCN was combined with multi-graph GCN and 3D CNN. The main contributions are as follows.

(1) The proposed multi-graph GCN has significant advantages to capture spatiotemporal and topological

correlations in a whole network. Three patterns (recent, daily, and weekly) are dealt with by the multi-graph GCN.

(2) The 3D CNN was used to innovatively integrate the inflow and outflow information as well as extract high-level correlations between three inflow/outflow patterns, and between stations located nearby and far away.

(3) This model was evaluated on the smart card data from the Beijing subway and obtained better performance than HA, ARIMA, LSTM, 2D CNN, ConvLSTM, ST-ResNet, ResLSTM, 3D CNN, and ST-GCN. Its outcomes were always the best no matter in which time interval, indicating strong robustness of this model.

(4) In terms of the RMSE, the performances under three time intervals have been improved by 9.402%, 7.756%, and 9.256%, respectively, showing promising results.

However, there are some limitations to this study. First, it is known that weather conditions have influences on passenger travels. We did not take weather factors into account. Second, there are significant randomness and low regularity for the passenger flow on weekends, which will affect prediction precision. In our study, we did not involve data on weekends. Therefore, researchers can make efforts to overcome these limitations in further studies. Moreover, model architecture can also be transplanted in other scenarios, such as taxi and bike-sharing systems in further studies.

## 5. Conflicts of Interest

The authors declare no conflict of interest.

## 6. Acknowledgments

This work was supported by the National Natural Science Foundation of China (grant numbers 71871027 and 51978044). The authors are grateful to financial support from the program of China Scholarships Council.